# Snakes partition their body to traverse large steps stably

Sean W. Gart[†], Thomas W. Mitchel, and Chen Li*

Department of Mechanical Engineering, Johns Hopkins University

3400 N. Charles St, 126 Hackerman Hall, Baltimore, Maryland 21218-2683, USA

*Email: [redacted]

## SUMMARY STATEMENT

Generalist snakes divide their body into sections, each using distinct movement patterns, to get over large step-like obstacles. Such body partitioning may be generally useful for diverse, complex 3-D terrain.

## ABSTRACT

Many snakes live in deserts, forests, and river valleys and traverse challenging 3-D terrain like rocks, felled trees, and rubble, with obstacles as large as themselves and variable surface properties. By contrast, apart from branch cantilevering, burrowing, swimming, and gliding, laboratory studies of snake locomotion focused on that on simple flat surfaces. Here, to begin to understand snake locomotion in complex 3-D terrain, we study how the variable kingsnake, a terrestrial generalist, traversed a large step of variable surface friction and step height (up to 30% snout-vent length). The snake traversed by partitioning its body into three sections with distinct functions. Body sections below and above the step oscillated laterally on horizontal surfaces for propulsion, while the body section in between cantilevered in a vertical plane to bridge the large height increase. As the animal progressed, these three sections traveled down its body, conforming overall body shape to the step. In addition, the snake adjusted the partitioned gait in response to increase in step height and decrease in surface friction, at the cost of reduced speed. As surface friction decreased, body movement below and above the step changed from a continuous lateral undulation with little slip to an intermittent oscillatory movement with much slip, and initial head lift-off became closer to the step. Given these adjustments, body partitioning allowed the snake to be always stable, even when initially cantilevering but before reaching the surface above. Such a partitioned gait may be generally useful for diverse, complex 3-D terrain.

[†]Present address: US Army Research Lab, Aberdeen Proving Ground, MD, USA 21005

## INTRODUCTION

Snakes are exceptionally versatile animals and can use their slender, highly articulated, near continuum bodies to move through almost any environment (Byrnes and Jayne, 2012; Gans, 1986; Goldman and Hu, 2010; Gray and Lissmann, 1950; Jayne, 1986; Lillywhite et al., 2000; Marvi et al., 2014; Munk, 2008; Socha, 2002). Many snakes live in deserts, forests, mountains, and coastal areas with felled trees, boulders, and branches, which present large 3-D obstacles comparable to their body size (Li et al., 2015). By contrast, with the exception of arboreal (Astley and Jayne, 2007a; Lillywhite et al., 2000) and burrowing (Sharpe et al., 2014) snakes, our understanding of terrestrial snake locomotion has been relatively limited to that on flat surfaces, whether they are level, sloped, granular, scattered with peg arrays, or confined between channels (Gray, 1946; Marvi and Hu, 2012; Hu et al., 2009; Jayne, 1986; Schiebel et al; Moon and Gans, 1998; Marvi et al., 2014, Astley et al., 2015).

Terrestrial snakes use four distinct locomotor gaits—lateral undulation, concertina, rectilinear, and sidewinding (Astley et al., 2015; Gans, 1962; Gans, 1986; Gray, 1946; Jayne, 1986; Marvi et al., 2014; Mosauer, 1932)—to move about depending on surface properties and geometric constraints of the environments. Most studies of snake locomotion on flat surfaces using these gaits observed nearly 2-D body deformation (no more than 10% body deformation out of the transverse plane, Table S1, Fig. S1, brown), which is unlikely to be effective in terrain with large 3-D obstacles relative to body size. Similarly, 2-D theoretical models developed for snake locomotion on flat surfaces (Alben, 2013; Guo and Mahadevan, 2008; Hu et al., 2009; Marvi and Hu, 2012; Marvi et al., 2013) do not directly apply to snake locomotion in the 3-D terrain common in nature.

In this study, we investigate terrestrial snakes traversing a large step obstacle (Gart et al., 2017) to begin to discover terradynamic principles (Li et al., 2013) of limbless locomotion in complex 3-D terrain (Li et al., 2015). We chose the variable kingsnake (*Lampropeltis mexicana*), a generalist found in diverse rocky habitats ranging from deserts to pine-oak forests (Hansen and Salmon, 2017), because it regularly traverses a variety of large step-like obstacles such as brush, boulders, and felled trees. Because these step-like obstacles have a broad range of size and surface properties, we varied step height and surface friction to test whether and how the snake changes its body movement in response to, and whether and how its performance is affected by, these terrain variations.

## MATERIALS AND METHODS

### Animals

We used three captive-bred juvenile variable kingsnakes (*Lampropeltis mexicana*). Snakes were housed in 60 × 20 cm containers on a 12:12 hour light:dark schedule at a temperature of 30 °C. We fed the snakes a diet of water and pinky mice. The snakes' snout-to-vent length (SVL) measured 34.6 ± 0.4 cm (mean ± s.d.) and full body length measured 39.6 ± 0.4 cm, and they weighed 19.7 ± 0.3 g. We measured snake length by digitizing dorsal view photos (Mendelson et al., 2017). To quantify body tapering, we measured the snake's cross-sectional width and height using calipers at 10 equally spaced (by approximately 0.3 cm) points from the neck to the vent (Fig. S2A, B).

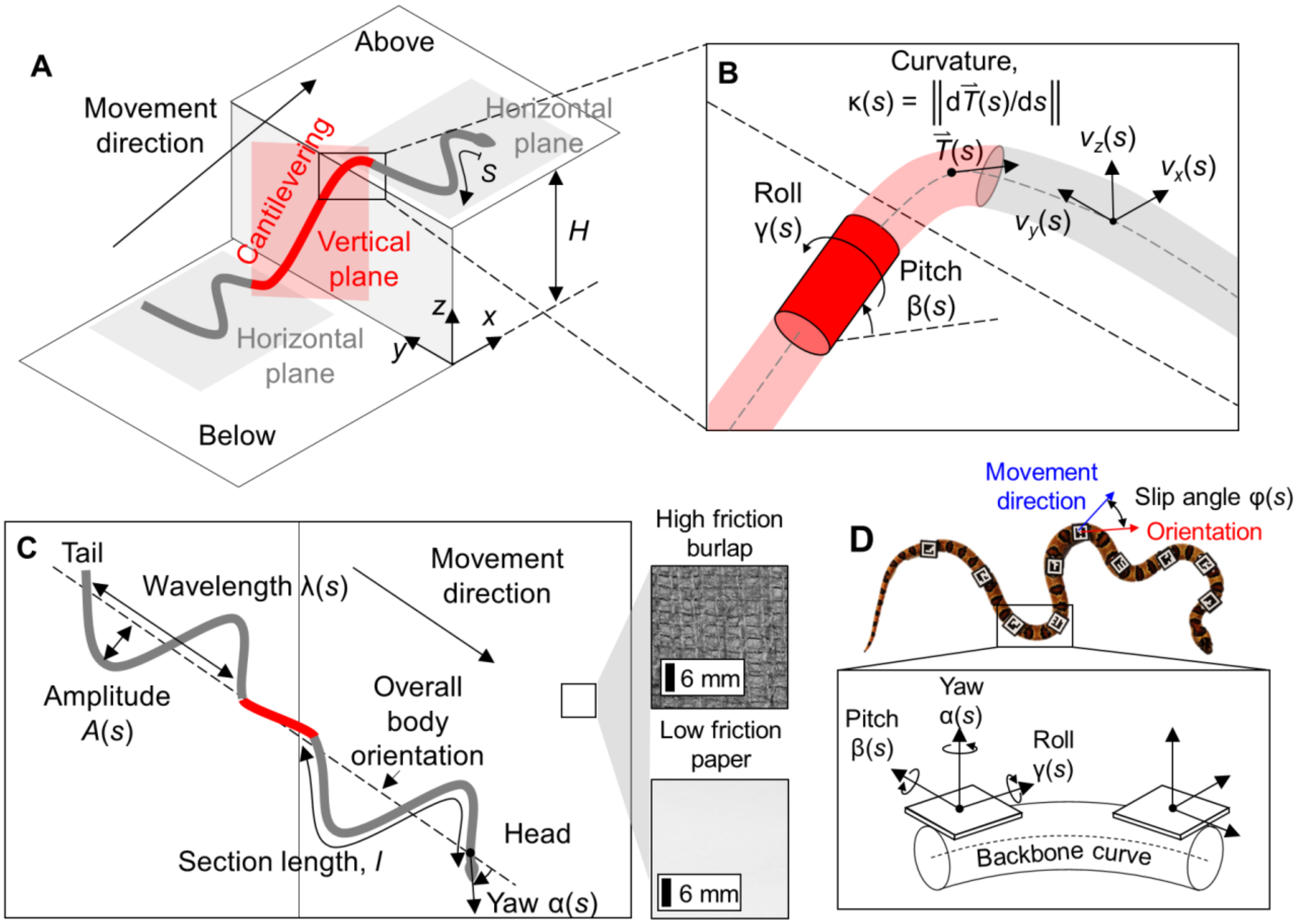

**Fig. 1. Schematic of partitioned gait and definition of kinematic variables.** (A) Oblique view schematic of snake traversing a large step by partitioning body into a cantilevering section (red) and two sections below and above step (gray). Red box is a vertical plane that the cantilevering body section moves in. Body coordinate, $s$, and step height, $H$, are defined. (B) A close-up schematic defining roll $\gamma$, pitch $\beta$, body tangent unit vector $\vec{T}$, curvature $\kappa$, and fore-aft $v_x$, lateral $v_y$, and vertical $v_z$ speeds, all of which are functions of body coordinate $s$. (C) Top view schematic showing definition of lateral oscillation wavelength $\lambda$ and amplitude $A$, section length $l$, and yaw $\alpha$. Dashed line shows overall body orientation in the horizontal plane. To test the effect of surface friction, we covered the step with either high friction burlap or low friction paper (insets). (D) A variable kingsnake with BEEtags (Crall et al., 2015) from the neck to the vent to measure 3-D position ($x$, $y$, $z$) and orientation ($\alpha$, $\beta$, $\gamma$). We used a mechanics-based model (Kim and Chirikjian, 2006) to interpolate between markers to obtain a backbone curve that describes the continuous

body 3-D position and orientation from the neck to the vent. Slip angle φ is defined as the angle between local forward orientation (red) and movement direction (blue) of a body segment (Sharpe et al., 2014).

**Step obstacle track**

We constructed a 120 cm long, 90 cm wide obstacle track using extruded T-slotted aluminum and acrylic sheets (McMaster-Carr, Elmhurst, IL, USA). The step spanned the entire width of the track. No sidewalls were used to prevent the snakes from using concave corners between the step and sidewalls to traverse. To study the effect of step height, we used two step heights, $H = 5$ cm ($\approx$ 15% SVL) and $H = 10$ cm ($\approx$ 30% SVL) (Fig. 1A). To study the effect of surface friction on snake locomotion, we covered the surface of the track either with a high friction burlap or a low friction paper (Pacon 4-ply railroad poster board, Appleton, WI, USA) (Fig. 1C, inset).

**Friction coefficient measurements**

We measured static friction coefficient between the snake body and the burlap and paper surfaces using three euthanized animals, with experimental protocols following (Hu et al., 2009). The animals were euthanized with the assistance of a veterinarian by intramuscular injection of ketamine (10-50 mg/kg) and medetomidine (0.1-0.15 mg/kg), followed by an overdose of a barbiturate into the coelomic cavity. The animal was monitored until a cessation of all cardiac and respiratory activity was confirmed. We laid euthanized snakes flat and straight on an inclined plane and increased its slope until the snakes began to slide, at which the angle of inclination $\phi$ was recorded. The coefficient of static friction $\mu$ was then estimated as $\mu = \tan\phi$.

We varied the forward direction of the straight snake body relative to the axis of rotation of the inclined plane to obtain static friction coefficient in the forward, transverse, and backward directions along the snake body (Hu et al., 2009). Each individual was tested three times for each surface and orientation treatment. For the high friction burlap surface, static friction coefficient in the forward, transverse, and backward directions were $\mu = 0.15 \pm 0.03$, $0.21 \pm 0.04$, and $0.49 \pm 0.17$ (mean ± s.d.), respectively. For the low friction paper surface, static friction coefficient in the forward, transverse, and backward directions were $\mu = 0.11 \pm 0.03$, $0.12 \pm 0.02$, and $0.17 \pm 0.03$ (mean ± s.d.), respectively.

**Locomotion experiment protocol**

We recorded snake locomotion using seven high speed cameras (Adimec, Eindhoven, Netherlands) with a resolution of 2592 × 2048 pixels at 100 frames $s^{-1}$. The experiment arena was illuminated by two 500 W halogen lamps and two LED light strips placed dorsally above the track. The lights heated the surface of the test area to 35 °C. To track the 3-D movements of the snake, we attached twelve to fourteen 1 × 1 cm BEEtags (Crall et al., 2015) equally spaced (≈ 1.5 cm) along the dorsal side of the body from neck to vent (Fig. 1D). We chose to use BEEtags because they provided both 3-D position and 3-D orientation information of the snake body at each tag location. This was important because we observed large changes in local body orientation during large step traversal.

We attached markers starting at the neck to avoid obscuring the snake's vision and ending at the vent because the tail was too narrow to reliably attach the markers. To attach the markers to the snake, we first attached each marker to a small, lightweight (1 g) 3-D printed mount. We then attached the mount to a piece of lightly adhesive tape (0.3 × 0.5 cm) using superglue and attached the tape onto the dorsal surface of the snake. The mount dorsally offset the marker by 0.3 cm from the body so that the snake body could bend dorsally and laterally without marker interference. We digitally moved the 3-D marker position ventrally (in the downward direction perpendicular to the marker plane orientation) by 0.3 cm plus local body radius to find the center of the body cross section of the snake below each marker, using body radius measurements to account for tapering of the body.

Snakes were kept in a container near the test area at a temperature between 25-30 °C prior to experiments. We placed snakes on the track one at a time for testing. During each trial, the snake was encouraged to traverse the step obstacle by light tapping on the tail and a shaded shelter at the end of the track. After each trial, we immediately removed the snake from the test area and placed it in the container to rest for 1-2 minutes. The snake did not remain on the track for more than one minute to avoid overheating.

**Discrete 3-D kinematics reconstruction using markers**

To calibrate the cameras for 3-D reconstruction, we made a 70 × 70 cm calibration grid out of Lego bricks (The Lego Group, Bilund, Denmark). We attached BEEtags (Crall et al., 2015) on all but the bottom sides of 3-D printed caps that we placed at the top of each landmark at the top of a Lego pillar to enable automatic tracking of each landmark's 2-D coordinates in each camera view for 3-D calibration. The calibration grid placement was carefully chosen to ensure that each camera captured at least 15 landmarks for reliable calibration. We obtained intrinsic (focal length, principal point, and pixel skew) and extrinsic (relative position and rotation) camera parameters using direct linear transformation (DLT) (Hedrick, 2008). After experiments, we exported videos into a custom MATLAB script to track the markers in each camera view using the BEEtag code (Crall et al., 2015). We then used a custom DLT script (Hedrick, 2008) to obtain 3-D position (fore-aft $x$, lateral $y$, and vertical $z$) and orientation (yaw $\alpha$, pitch $\beta$, and roll $\gamma$). We calculated Euler angles using the Z-Y'-X" Tait-Bryan angle convention, where ' and " denotes the coordinate frame after the first and second rotations, respectively. The $x$, $y$, $z$ axes of the lab frame align with the forward, lateral, and vertical directions relative to the step. Changes in roll along the body measure how much the body is twisting. Although snake vertebrae typically can only twist about two degrees per vertebra (Jurestovsky and Astley, 2019), over many vertebrae there could be large twisting. Rolling of the skin relative to the vertebrae and rib motion can also result in nominal twisting (Henry Astley, personal communication).

**Continuous body 3-D kinematics interpolation**

To determine when and which part of the body contacted the 3-D terrain, we obtained a continuous body 3-D description of the snake during locomotion. In preliminary data analyses, we found that a discrete description of the body movement using only the marker data was not sufficient to describe how the snake interacted with the step (and complex 3-D terrain in general). Even with a high density of greater than 10 BEEtag markers distributed over 30 cm of snake body, stick figures of the body shape connecting markers often penetrated the corner of the step.

To address this issue, we obtained continuous body 3-D kinematics by interpolating the sections of body between adjacent markers using a mechanics-based method (Chirikjian and Burdick, 1995; Kim and Chirikjian, 2006). This method approximates the snake body as a passive elastic rod (Cheng et al., 1998), divides the body section between adjacent markers into many small segments, and it obtains the 3-D position and orientation of each body segment by using marker position and orientation as end constraints (Fig. 1D) and minimizing the bending energy using elastic rod theory (Kirchhoff, 1859). We interpolated every other BEEtag as end constraints because using every BEEtag over-constrained the interpolation due to small but finite errors in marker placement (the marker inevitably had a small yaw and roll offset from the perfect local forward and upright directions of the body) which, when interpolated over a shorter marker separation, led to poorer interpolation results. Our ongoing experiments (Mitchel et al., 2018) showed that, despite such a drastic over-simplification (approximating snake body as an elastic rod), for the small marker separation (~5 cm) that we used for interpolation, this method well approximated the snake body's midline positions with a small error of $1.1 \pm 0.7$ mm ($10 \pm 6$ % of the maximum body radius) (mean ± s.d.) over the entire length of the body from neck to vent. In addition, this method does not assume that the snake body shape is a linear superposition of planar shape functions, a common assumption in methods for approximating 2-D body deformation during locomotion on flat surfaces (Gong et al., 2015; Sharpe et al., 2014; Shen et al., 2012). For the remainder of the paper, we refer to the interpolated midline of the snake (not including the head and the tail) as the backbone curve (Fig. 1D).

Using the backbone curve, we then reconstructed the surface of the snake body by expanding the backbone curve radially outward by the local body geometric radius ($(\text{width} \times \text{height})^{1/2}$), using body radius measurements to account for tapering of the body (Fig. S2A, B). Although the pre-cloacal vertebrae of the snakes that we tested numbered between 208 and 226 (from counting the number of ventral scales from snout to vent (Voris, 1975), considering that the snake's body surface, which interacted with the environment, is nearly continuous, we used 2000 segments for backbone curve interpolation to obtain a near continuous surface of the body to better quantify body-step interaction. Thus, a segment in our interpolation does not represent one vertebra.

**Performance analysis**

To quantify the snake's large step traversal performance, we measured traversal speed and traversal time. Traversal was defined from when the neck (the most anterior marker) first lifted off of the surface below the step to when the vent (the most posterior marker) reached the surface above the step. Traversal speed was the average center of mass speed $v_{\mathrm{CoM}}$ during this process. Traversal time was the duration of this process. We measured how intermittent the snake's movement was using coefficient of variation of $v_{\mathrm{CoM}}$, the ratio of the standard deviation of $v_{\mathrm{CoM}}$ to the mean $v_{\mathrm{CoM}}$ of the entire trial (Jayne, 1986).

Because part of the snake's body cantilevered (Jayne, 2012) with a large body pitch to bridge the large height increase of the step (see **Partitioning of body into three sections** in **RESULTS**), we used a body pitch threshold to automatically separate which part of the snake body cantilevered (Fig. 1A). We considered body segments to be cantilevering if the local pitch of the body segment was above 25°. It is likely that the body began cantilevering at pitch angles as small as a few degrees; however, we chose this large threshold to remove false identifications of cantilevering seen at lower thresholds due to measurement and interpolation noise. Because the small portion of the body where surface lift-off and touch-down began had large out-of-transverse-plane curvatures, this larger threshold only slightly under-estimated the length of the body that lifted off of the surface and cantilevered (by 6% SVL) and slightly over-estimated the length of the body section in contact with the surfaces below and above the step (by 6% SVL). Examination of spatiotemporal profiles of vertical speed (Fig. 5A-D, iii) showed that the threshold chosen generated good separation of the cantilevering section, despite this under-estimation.

**Body partitioning analysis**

Because we observed that the snake's body was partitioned in three sections with distinct movement patterns (see **Partitioning of body into three sections** in **RESULTS**), we fit each of the three body sections to a plane to quantify its shape and movement (Fig. 1A).

For the two body sections below and above the step oscillated laterally on the horizontal surfaces (see **Oscillation in the horizontal planes on high/low friction steps** in **RESULTS**), we fit their respective movement by a traveling wave within a horizontal plane (Fig. 1A, gray), whose traveling direction was aligned with, but opposite to, the overall direction of the body, defined by the direction of the best linear fit of the body projection into the horizontal plane, relative to the forward direction ($+x$ axis) (Fig. 1C). For the sections below and above the step, wave amplitude $A$ is half the distance in the direction perpendicular to overall body orientation between alternating peaks (a peak on the left/right side followed by another peak on the right/left side). Wave frequency $f$ is the inverse of the duration between two consecutive maximal lateral displacements. Wavelength $\lambda$ is the distance between wave peaks along the overall body orientation. Wavenumber $\sigma$ is the length (along the body midline) of the body section normalized by wavelength. We note that the traveling waves were highly variable and far from perfectly sinusoidal and that these wave properties only characterize the general pattern of movement.

For the cantilevering body section whose deformation was nearly planar and which was nearly stationary relative to the step during traversal (see **Cantilevering in the vertical plane** in **RESULTS**), we fit its movement to a vertical plane (Fig. 1A, red), whose orientation in the horizontal plane varied with time. We used its shape projected into the fit plane, averaged across all frames for each trial, to represent the cantilevering body section's shape.

To quantify how well the body sections above and below the step lied within the horizontal plane and how well the cantilevering body section lied within a vertical plane, we calculated the in-plane component of each body section as the ratio of its length projected into the plane to its total length. The out-of-plane component is then one minus the in-plane component.

**Kinematics analysis**

We calculated fore-aft speed $v_x$, lateral speed $v_y$, and vertical speed $v_z$, velocity magnitude $v$, yaw α, pitch β, and roll magnitude |γ| of each infinitesimal body segment as a function of the section's body coordinate $s$, the cumulative length along the body from the neck (the most anterior marker) (Fig. 1A, B). Because horizontal-plane overall body orientation varied by up to ± 30°, to more clearly show lateral deformation of each body segment relative to the overall body orientation, for each trial, we calculated local body yaw α relative to horizontal-plane overall body orientation at each instance (Fig. 1C, dashed line). To measure how straight (or curved) the body is locally, we calculated local body curvature, $\kappa(s) = \|\mathrm{d}\vec{T}(s)/\mathrm{d}s\|$, the magnitude of the spatial derivative of the body tangent unit vector, $\vec{T}(s)$, as a function of the body coordinate $s$. We verified that this is equivalent to calculating curvature by the inverse of local radius of curvature along the body (less than 5% difference between the two methods).

To quantify how straight the snake moved on the horizontal surfaces during step traversal, we calculated tortuosity of the center of mass trajectory, τ, defined as the ratio of the total length along the center of mass trajectory from the start to the end of each trial to the distance between the starting and ending positions of the center of mass of the trial (Jayaram and Full, 2016). A lower tortuosity means a straighter trajectory, with a minimal tortuosity of one for a perfectly straight trajectory. A higher tortuosity means a more meandering trajectory.

To measure the amount of slip that hinders forward movement of the snake, we calculated the slip angle, φ, the angle between the local movement direction and the local forward orientation of each body segment (Sharpe et al., 2014). Slip angle is high when the body slips laterally or backwards relative to its orientation; slip angle is low when the body slips little laterally or backwards, such that each segment of the body more closely follows the previous one as if the animal were "moving in a tube" (Gray, 1951; Hu et al., 2009; Sharpe et al., 2014). For the cantilevering body section, slip angle measurements only reflected how much the body deviated from the "tube" and not actual slipping, because there was no surface contact against which to slip.

**Static stability analysis**

Our continuous body 3-D description allowed us to examine static stability of the snake during traversal. We performed a static stability region analysis (Ting et al., 1994) for three stages of traversal: before cantilevering, during cantilevering but before reaching the surface above the step, and after reaching the surface above the step (see **Static stability** in **RESULTS**). We calculated body center of mass position and the maximal convex region in the horizontal plane (Ting et al., 1994) formed by all the body segments in contact with both the horizontal surfaces below and above the step (i.e., convex hull (Preparata and Hong, 1977)). We assumed that the entirety of body sections below and above the step are in contact with the horizontal surfaces. This is not necessarily true because snakes can slightly lift portions of the body during locomotion on horizontal surfaces (by a few mm vertically for similarly sized corn snakes) (Hu et al., 2009). However, although such this may reduce the size of the stability region, it is likely that snakes can achieve nearly as much stability as if there is no lifting, because the slightly lifted body can readily regain ground contact. The body was statically stable if the center of mass projection into the horizontal planes fell inside this stability region and unstable otherwise. When the snake was in a stable configuration, we measured stability margin in the pitching and rolling directions, defined as the minimal horizontal distance (within the $x$-$y$ plane) of the center of mass projection to the boundary of the stability region perpendicular to and parallel to the horizontal-plane overall body orientation, respectively.

Because we could not interpolate the body shape beyond the most posterior marker above the vent, this method did not directly account for possible ground contact by the tail below the step once the most posterior marker lifted off of the surface. To approximately account for this, we measured the length of the tail beyond the most posterior marker, projected this length along the posterior direction of the most posterior marker, and determined whether or not the projected "straight tail" intersected the surface below the step. If it did, we included the intersection point in the maximal convex region (see **Static stability** in **RESULTS**). Video observation showed that this slightly under-estimated the size of the stability region because the tail always curled dorsally before lifting off. We did not make this correction for the head because it only slightly extended forward beyond the first tag at the neck (approximately 1 cm or 3% SVL) and excluding it only resulted in a small under-estimation of the size of the stability region.

**Statistics**

We performed experiments with three snakes ($N$ = 3) for each of the two step heights and each of the two friction treatments. We randomized the testing sequence for different treatments for each snake and tested the animal until it traversed 10 times for each treatment, resulting in a total of $n$ = 120 successful trials. We accepted trials if the entire snake body reached the surface above the step and remained in view of at least two cameras during the entire duration of traversal so that 3-D kinematics could be obtained. Trials were discarded if the snake failed to traverse or moved out of six or all of the seven camera views. We allowed the snake to move at its own chosen speed during each trial.

To compare measurements across treatments, for each trial, we first averaged $v_x$, $v_y$, $v_z$, $\alpha$, $\beta$, $\gamma$, $\varphi$, $\kappa$, $\lambda$, $A$, $f$, and $\sigma$ spatially across each of the three body sections (below, cantilevering, and above) for each video frame. We then averaged these section averages temporally to obtain the means for each body section for the trial to obtain the values used in ANOVAs. Finally, we calculated the means and standard deviation (s.d.) of the spatiotemporally averaged means using all trials from all individuals for each treatment. For measurements relating to the entire animal body including $v$, traversal time, cantilever section length, velocity intermittency, and $\tau$, we averaged spatially across the entire reconstructed body.

To determine which variables affected traversal performance, we used a fully crossed mixed-effects ANOVA, with body section, step height, and surface treatment as fixed, crossed factors and individual used as a random, crossed factor (to account for individual variation). Details of statistical test results using ANOVAs are shown in Tables S2, S3, and S4. We used Tukey's honestly significant difference (Tukey's HSD) test for post-hoc analysis.

All data reported with variation are means ± 1 standard deviation (s.d.).

## RESULTS

### Traversal performance

The animal's step traversal performance decreased with step height and increased with surface friction. For both the 15% SVL and the 30% SVL step, traversal time more than doubled as surface friction decreased. For both low and high friction surfaces, traversal time increased by about 50% as step height increased from 15% SVL to 30% SVL (Fig. 2A; $P < 0.0001$, ANOVA, Table S2). In addition, traversal speed (magnitude of center of mass velocity during traversal) decreased by about 30% as surface friction decreased (Fig. 2B; $P < 0.0001$, ANOVA, Table S2). Traversal speed was slightly lower on the higher 30% SVL step, although this difference was not statistically significant (Fig. 2B; $P > 0.05$, ANOVA, Table S2). Throughout traversal, the snake's body acceleration was small (mean ± s.d. = 0.02 ± 0.01 m $s^{-2}$, only 0.2 ± 0.1 % of gravitational acceleration), meaning that the animal moved quasi-statically.

In failed trials, the snakes either did not attempt to traverse the step or tried to move around it. In preliminary experiments, after 90 attempted trials, no animal was able to traverse an even higher 45% SVL step covered with either high friction burlap or low friction paper. In all experiments, we did not observe any snake toppling off the step.

### Partitioning of body into three sections

Regardless of changes in step height or surface friction, the snake always traversed the step by partitioning its body into three sections with distinct movement patterns (Fig. 1A, Fig. 2C-F, Movie 1), with large body deformation out of the transverse plane (Fig. S1, red). The posterior body section below the step and the anterior body section above the step both remained in contact with the horizontal surfaces and oscillated laterally, with a wave-like pattern traveling down the body to propel the animal forward (Fig. 1A, gray). The lateral oscillation was highly variable and far from perfectly sinusoidal. To bridge the large height increase of the step, the body section in between cantilevered (Fig. 1A, red). See **Body partitioning analysis** in **MATERIALS AND METHODS** for quantitative definition of body sections.

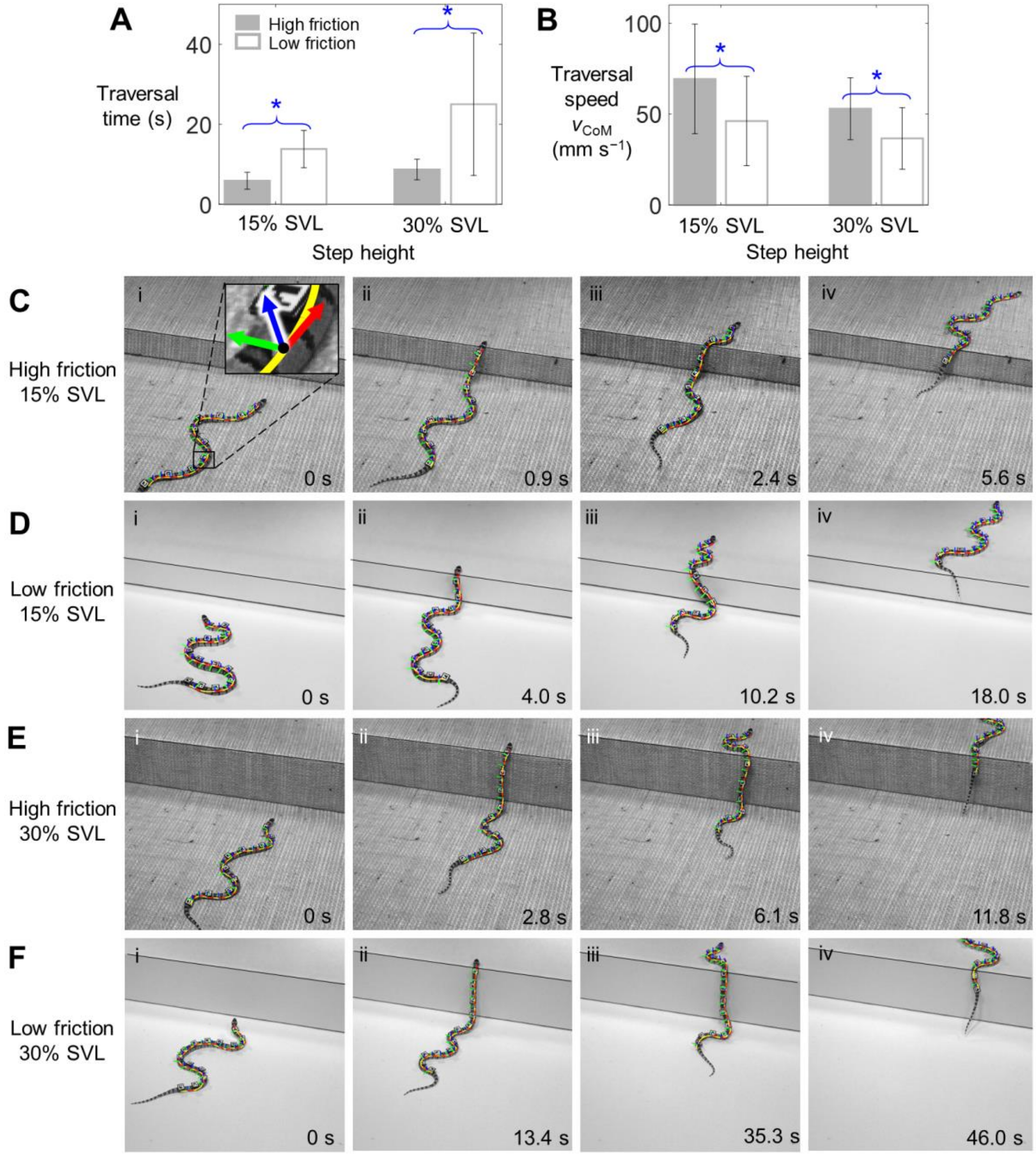

**Fig. 2. Traversal performance and representative snapshots.** (A) Traversal time as a function of step height. (B) Speed (magnitude of total velocity) during traversal as a function of step height. In (A) and (B), filled and open bars are high and low friction, respectively. Error bars show ± 1 s.d. Blue brackets and asterisks represent a statistically significant difference between surface friction ($P < 0.05$, ANOVA, Table S2). (C-F) Representative oblique view snapshots of snake traversing a high friction 15% SVL step (C), a low friction 15% SVL step (D), a high friction 30% SVL step (E), and a low friction 30% SVL step (F). Snapshots show (i) prior to cantilevering, (ii) the head reaching the surface above the step, (iii) the snake's body partitioned into three sections, (iv) the tail lifted off of the surface below the step. Yellow curve is the backbone curve (Fig. 1D). Red (forward), blue (dorsal), and green (lateral) arrows show body segments' local reference frames (see inset in (C), first frame). See Movie 1 for videos of representative trials of each treatment.

In addition, as the snake progressed forward and upward onto the step, the three body sections travelled down each body segment (Fig. 2C-F, Movie 1). First, as the snake laterally oscillated on the horizontal surface below the step and progressed towards it (Fig. 2C-F, i), each body segment consecutively lifted off of the surface and cantilevered upward and forward (Fig. 2C-F, ii). Then, after reaching the surface above the step, each body segment regained surface contact and resumed lateral oscillation on the horizontal surface (Fig. 2C-F, iii). Thus, the body section below continued to shorten and eventually disappeared and the body section above continued to lengthen, while the cantilevering body section remained nearly constant in length (Fig. 2C-F, ii-iv).

The distinct movement patterns of the snake's three body sections were further reflected by differences in speeds and orientations between sections (Fig. S2C-G). Compared to the body sections below and above the step, the cantilevering body section had a higher upward speed for all step height and surface friction treatments (Fig. S3E, red asterisks; $P < 0.0001$, ANOVA, Tukey HSD, d.f. = 1, 2, Table S3) and lower forward and lateral speeds for all but the high friction 15% SVL step (Fig. S3C, D; $P < 0.05$, ANOVA, Tukey HSD, d.f. = 1, 2, Table S3). In addition, for all step height and surface friction treatments, the cantilevering body section pitched and rolled more than the body sections below and above the step (Fig. S3F, G, red asterisks; $P < 0.0001$, ANOVA, Tukey HSD, d.f. = 1, 2, Table S3).

**Planar movement of each body section**

Although the snake's body overall displayed large deformation in three dimensions, body movement within each of the three sections was nearly two-dimensional. Both the body section below and above the step moved almost entirely within a horizontal plane on the surface (in-plane component = 98 ± 1 % and 99 ± 1 %, respectively, Fig. 3A). By contrast, the cantilevering body section moved almost entirely within a vertical plane (in-plane component = 94 ± 4 %, Fig. 3A). Note that this vertical plane of the cantilevering body section did not always align with the overall body orientation in the horizontal plane as its orientation varied with time (± 30°). These horizontal planar and vertical movements are clearly seen in front view projection in the insets of Movie 1.

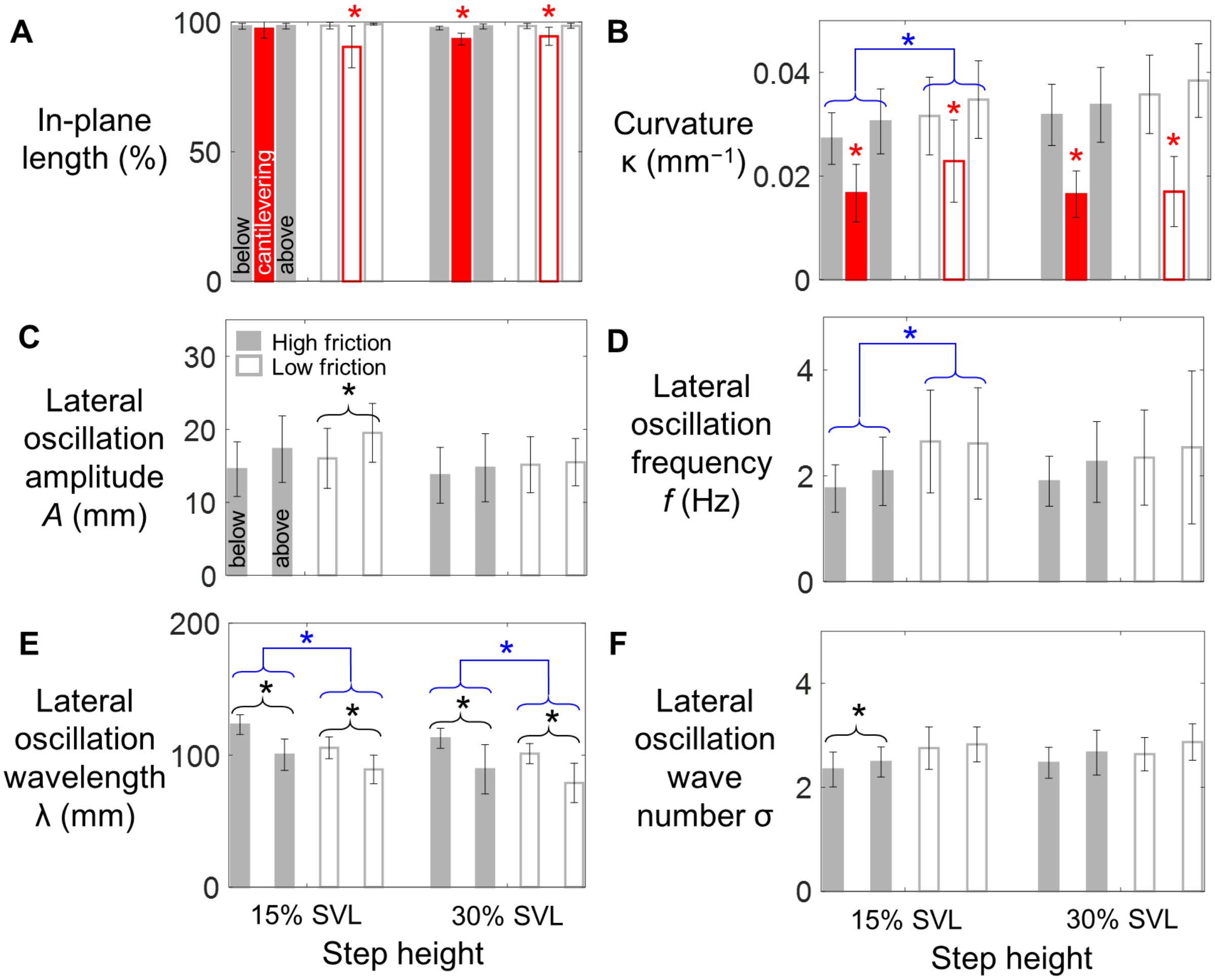

**Fig. 3. Body partitioning into three planes.** (A-F) Percentage of in-plane length (A), curvature (B), lateral oscillation amplitude (C), lateral oscillation frequency (D), lateral oscillation wavelength (E), and lateral oscillation wavenumber (F) of each body section in a horizontal plane (sections below and above) or a vertical plane (cantilevering section) (see Fig. 1A), as a function of step height and surface friction. In (A-F), filled and open bars are for high and low friction treatments, respectively. For each treatment, two gray bars are for body sections below (left) and above (right) step, and red bar in between is for cantilevering body section. Red bar is not shown in (C-F) because lateral oscillation does not occur for cantilevering body section. Error bars show ± 1 s.d. Brackets and/or asterisks represent statistically significant differences between body sections below and above the step (black) and between surface friction treatments (blue); red asterisks indicate that cantilevering section differs from body sections below and above the step ($P < 0.05$, ANOVA, Tables S3, 4). Connected brackets represent a significant difference across treatments for all body sections.

For all step height and surface friction treatments, the cantilevering body section was straighter (with a smaller curvature) than the body sections below and above the step (Fig. 3B, red asterisks; $P < 0.0001$, ANOVA, Tukey HSD, d.f. = 1, 2, Table S3). In addition, for all step height and surface friction treatments, lateral oscillation wavelength was larger for the body section below than that above the step (Fig. 3E, black brackets and asterisks; $P < 0.0001$, ANOVA, Table S4), while lateral oscillation frequency did not differ between the body section below and that above the step (Fig. 3D; $P > 0.05$, ANOVA, Table S4). In addition, lateral oscillation amplitude, frequency, and wavenumber did not differ across step height and surface friction treatments (Fig. 3C, D, F; $P > 0.05$, ANOVA, Table S4). For both the 15% SVL and 30% SVL step, lateral oscillation wavelength increased with surface friction (Fig. 3E, blue brackets and asterisks; $P = 0.034$, ANOVA, Table S4).

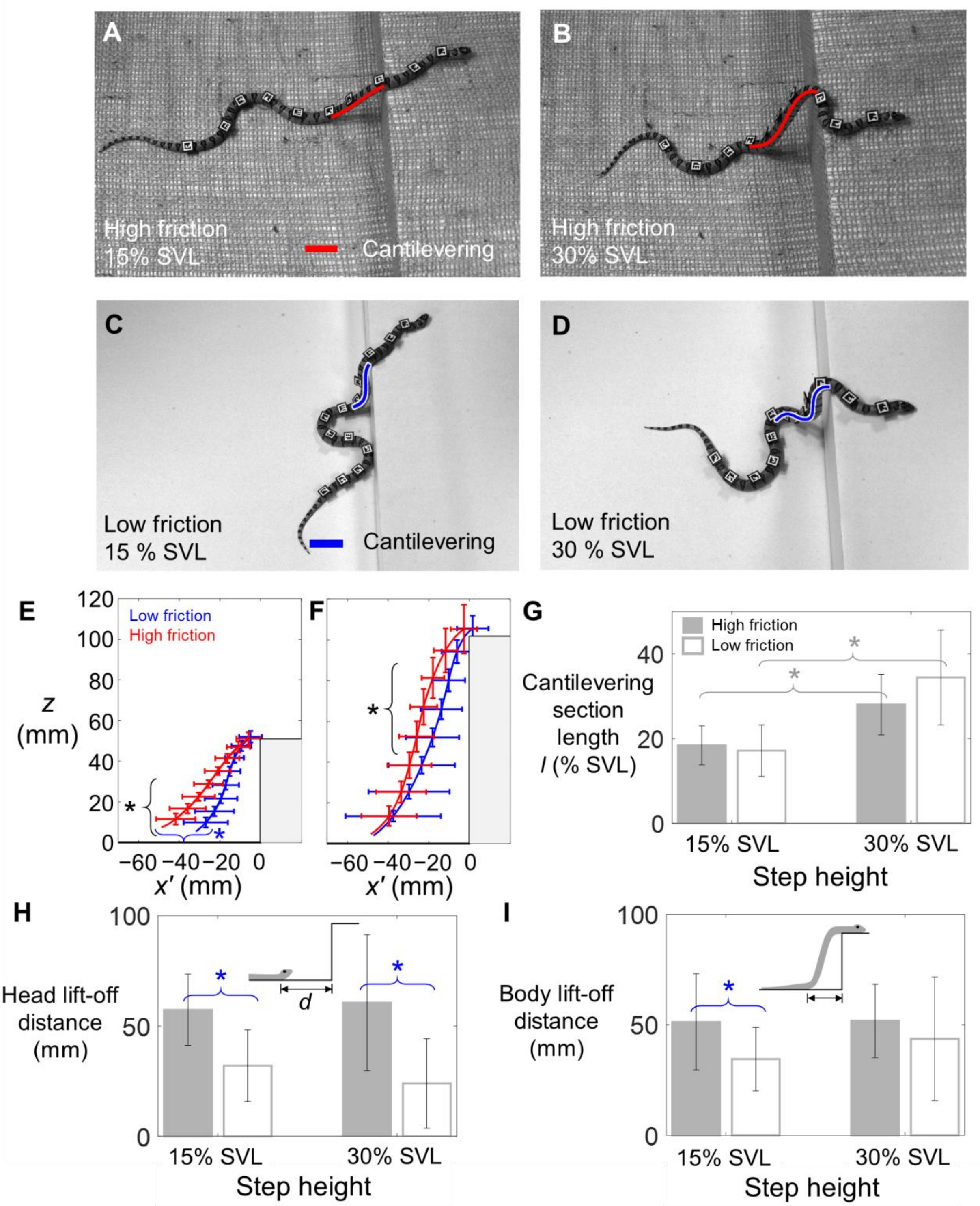

**Fig. 4. Cantilevering kinematics.** (A-D) Representative side view snapshots (looking slightly downward) of snake traversing a high friction 15% SVL step (A), a high friction 30% SVL step (B), a low friction 15% SVL step (C), and a low friction 30% SVL step (D). Cantilevering section is shown in red for high

friction treatments (A, B) and in blue for low friction treatments (C, D). (E, F) Cantilevering section shape in the vertical plane (aligned with time dependent overall body orientation) on the 15% SVL (E) and 30% SVL step (F) with a high (red) and low (blue) friction surface. $x'$ is the distance to the step in the vertical plane aligned with overall body orientation in the horizontal plane. Black brackets and asterisks represent the portion of cantilevering section whose horizontal distances to the step differ between low and high step friction treatments ($P < 0.05$, ANOVA. (G) Cantilevering body section length as a function of step height. (H) Initial head lift-off distance and (I) body lift-off distance during cantilevering as a function of step height. Inset shows definition of head and body lift-off distances. In (G-I), filled and open bars are for high and low friction treatments. Brackets and/or asterisks represent statistically significant differences between step height treatments (gray) and between surface friction treatments (blue) ($P < 0.05$, ANOVA, Table S2). Blue bracket and asterisk in (I) correspond with those in (E). Error bars show ± 1 s.d.

**Cantilevering in the vertical plane**

For all step height and surface friction treatments, as the snake's cantilevering body section traveled down the body, it maintained a relatively constant shape in the vertical plane (Fig. 4E, F), which resembled a shallow S-shape (Fig. 4A-F) similar to cantilevering body sections of snakes during arboreal locomotion (Byrnes and Jayne, 2012; Jayne and Riley, 2007). For both low and high friction steps, the length of the cantilevering section increased with step height (Fig. 4G, gray bracket and asterisk; $P = 0.005$, ANOVA, Table S2). Regardless of step height, the snake initially lifted its head to start cantilevering when it was farther from the step on the high friction step (59 ± 23 mm) than on the low friction step (28 ± 18 mm) (Fig. 4H, blue brackets and asterisks; $P < 0.05$, ANOVA, Table S2). However, after the initial head lift-off, body lift-off during cantilevering was farther from the step on the high friction step (51 ± 19 mm) than on the lower friction step (39 ± 21 mm) only for the lower 15% SVL step (Fig. 4I, blue bracket and asterisk). For both step heights throughout traversal, around 40% of the cantilevering section was closer to the step on a low friction step than it was to on a high friction step (Fig. 4E, F, black brackets and asterisks; $P < 0.05$, ANOVA.

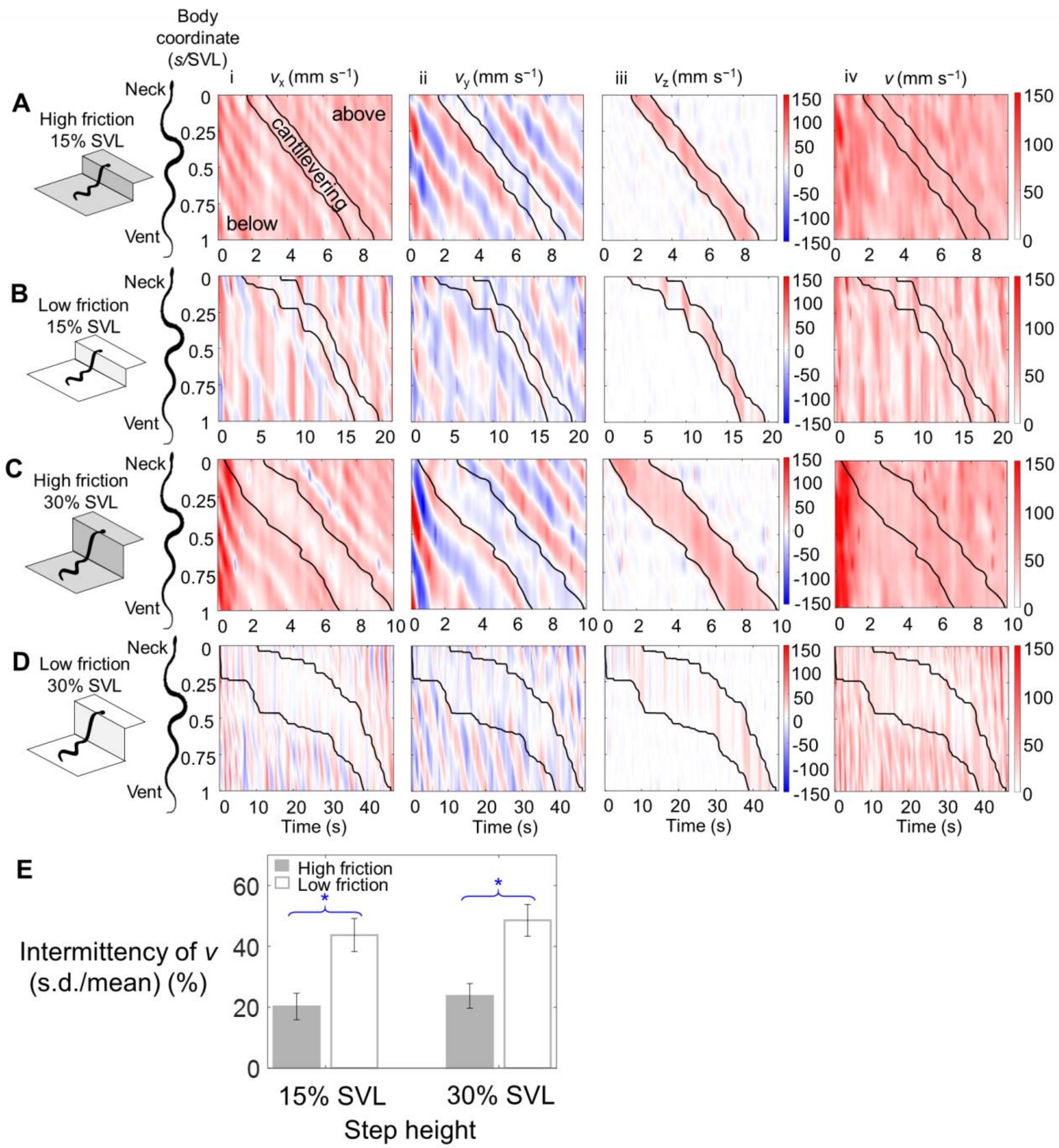

**Fig. 5. Body velocities.** (A-D) Representative spatiotemporal profiles of fore-aft (i), lateral (ii), vertical (iii), and total speed (iv) as a function of body coordinate and time for a high friction 15% SVL step (A), a low friction 15% SVL step (B), a high friction 30% SVL step (C), and a low friction 30% SVL step (D). The section between black curves is the cantilevering body section. Note the different time scale for between treatments. See Fig. 1A for definition of the three body sections. (E) Intermittency of velocity

(standard deviation relative to mean) as a function of step height. Filled and open bars are for high and low friction treatments, respectively. Error bars show ± 1 s.d. Blue brackets and asterisks represent a statistically significant difference between surface friction treatments ($P < 0.05$, ANOVA, Table S3).

**Oscillation in the horizontal planes on high friction steps**

Although the snake always used the partitioned gait to traverse the step, its body movement patterns changed in response to changes in step height and surface friction.

To traverse a high friction step, the snake's body undulated laterally both below and above the step, with an oscillatory wave continuously traveling down each section (Movie 1, part 1, 2), propelling the snake at a forward speed (Fig. 5A, i) and a total speed (Fig. 5A, C, iv) that were relatively uniform both spatially and temporally. Continuous lateral undulation of the body sections below and above the step was further evidenced by relatively uniform bands of alternating positive and negative lateral speeds (Fig. 5A, C, ii) and alternating positive and negative body yaw (Fig. 6A, C, i) traveling down the body. The rest of the body followed the head as if the entire body moved in a tube (Fig. 7A, B) with little slipping (Fig. 7E, solid bars). Although small, slipping did increase as step height increased (Fig. 7A, B; Fig. 7E, gray brackets and asterisks, filled bars; $P = 0.021$, ANOVA, Table S3). The continuous lateral undulation with little slip resulted in relatively straight center of mass trajectories in the horizontal planes (Fig. 7G-J, red solid curves), with a tortuosity only slightly larger than 1, which is for a perfectly straight trajectory (Fig. 7F, filled bars).

The straightened cantilevering body section with a relatively constant shape also continuously travelled down the body, as reflected by relatively uniform bands of high vertical speed (Fig. 5A, C, iii) and high body pitch (Fig. 6A, C, ii) traveling down the body, both with a nearly constant size along the body coordinate and a nearly constant slope, showing that the section length and speed traveling down the body were nearly constant. For both surface friction treatments, pitch and roll (magnitude considering lateral symmetry) of the cantilevering body section both increased with step height (Fig. S3F, G, gray brackets; $P < 0.0001$, ANOVA, Table S3). In addition, on the lower 15% SVL step, pitch and roll of the cantilevering section reduced with surface friction (Fig. S3F, G, blue bracket; $P < 0.0001$, ANOVA, Table S3).

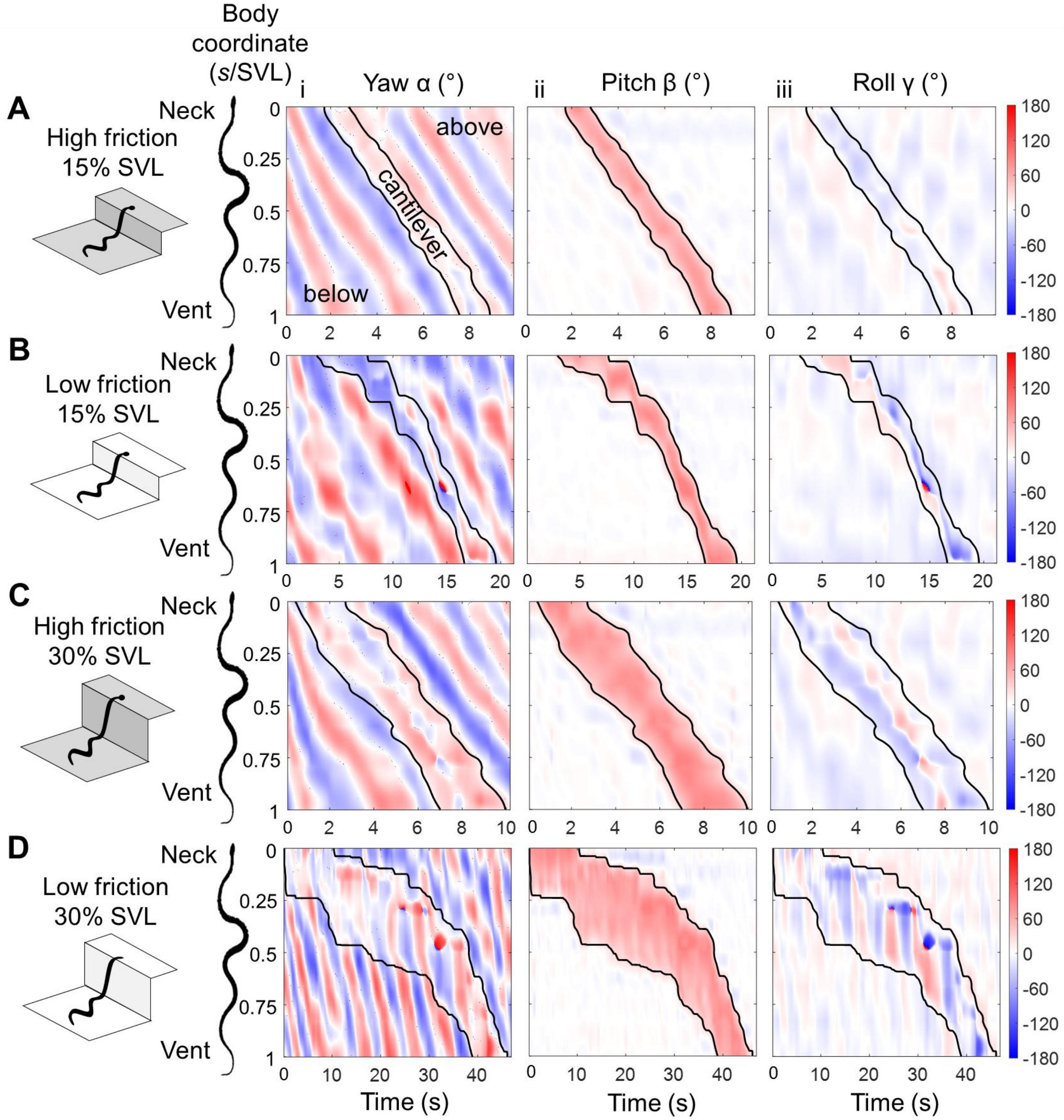

**Fig. 6. Local body orientations.** (A-D) Representative spatiotemporal profiles of body yaw (i), pitch (ii), and roll (iii) as a function of body coordinate and time for a high friction 15% SVL step (A), a low friction 15% SVL step (B), a high friction 30% SVL step (C), and a low friction 30% SVL step (D). The section

between black curves is the cantilevering section of the body. Note the different time scale between treatments. See Fig. 1A for definition of the three body sections.

**Oscillation in the horizontal planes on low friction steps**

On a low friction step, for both step heights, the snake moved more intermittently below and above the step (Movie 1, part 3, 4) and slipped more than on a high friction step (Fig. 7C, D, E, blue brackets and asterisks; $P = 0.006$, ANOVA, Table S2), with a more spatially and temporally variable forward speed (Fig. 5B, D, i), lateral speed (Fig. 5B, D, ii), and total speed (Fig. 5B, D, iv). The degree of intermittency, measured by the standard deviation of total speed relative to its mean (Jayne, 1986), was higher on a low friction step than on a high friction step for both step heights (Fig. 5E, blue brackets and asterisks; $P < 0.0001$, ANOVA, Table S2). Slipping also increased with step height (Fig. 7C, D; Fig. 7E, empty bars, gray brackets and asterisks; $P = 0.021$, ANOVA, Table S3). The cantilevering section traveled down the body more intermittently, as reflected by less uniform bands of high vertical speed (Fig. 5B, D, iii) and high body pitch (Fig. 6B, D, ii) traveling down the body.

Because of this intermittent movement, for both step heights, average forward and vertical speed were lower on the low friction step than on the high friction step (Fig. S3C, E; $P < 0.0001$, ANOVA, Table S3). In addition, the snake's center of mass trajectory in the horizontal plane was also visually less straight on a low friction step (Fig. 7G-J, blue dashed curves), with a larger tortuosity than on high friction steps (Fig. 7F, empty bars), although the difference in tortuosity was not statistically significant ($P = 0.0704$, ANOVA, Table S2). Pitch and roll (magnitude considering lateral symmetry) of the cantilevering body section also both increased with step height on the low friction steps (Fig. S3F, G, gray brackets and asterisks; $P < 0.05$, ANOVA, Table S3). On the 15% SVL step, they also both decreased with surface friction (Fig. S3F, G, gray brackets and asterisks; $P < 0.05$, ANOVA, Table S3).

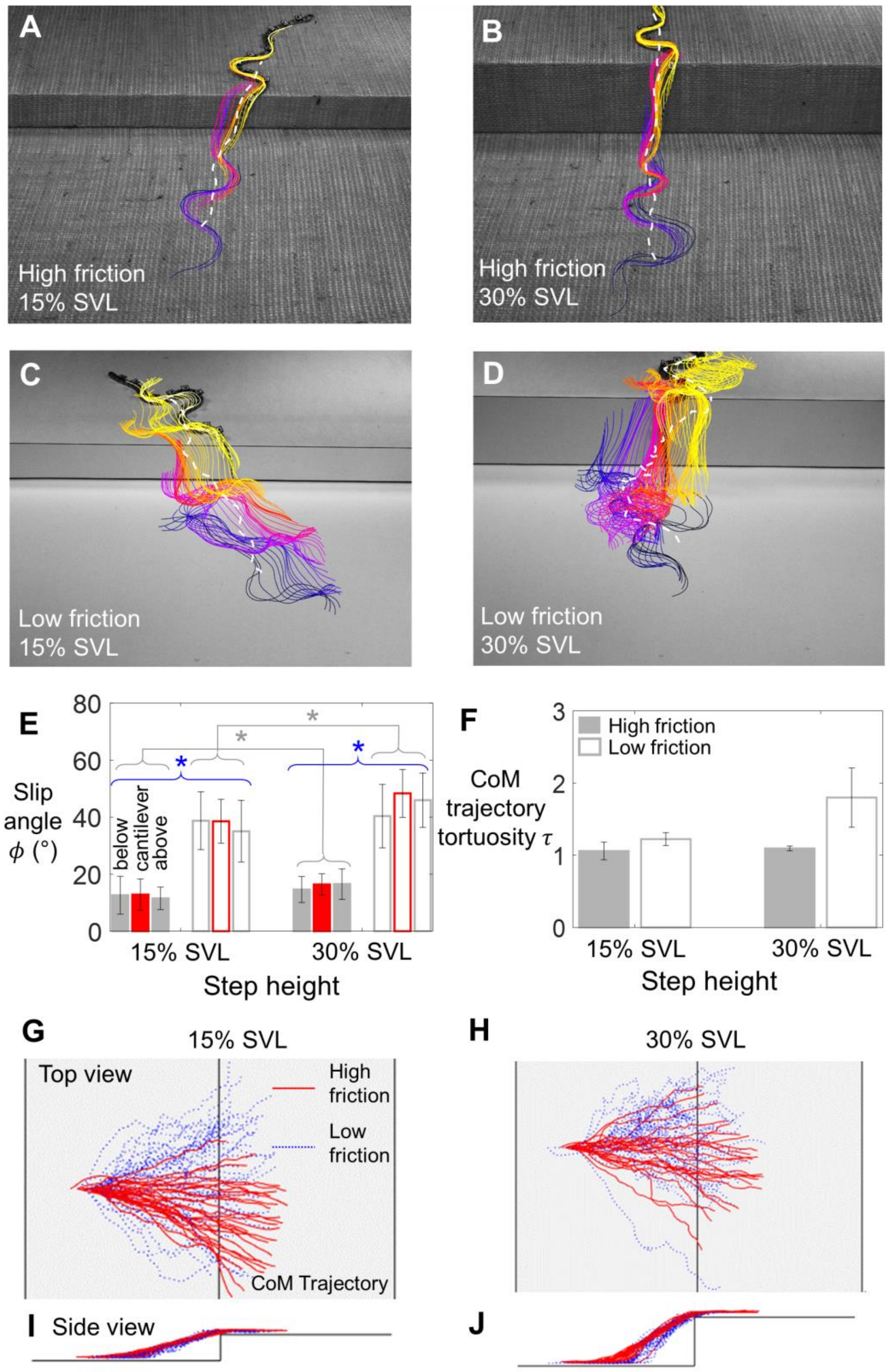

**Fig. 7. Slipping and center of mass trajectory.** (A-D) Representative rear view (looking slightly downward) snapshots of snake with backbone curve overlaid at different time instances during traversal of a high friction 15% SVL step (A), a high friction 30% SVL step (B), a low friction 15% SVL step (C), and a low friction 30% SVL step (D). Backbone curve color changes from dark blue to light yellow with elapse of time from start to end of traversal. White dashed curve is center of mass trajectory. (E) Slip angle $\varphi$ as a function of step height. For each treatment, two gray bars are for body section below (left) and above (right) step, and red bar in between is for cantilevering body section. For cantilevering section, slip angle merely measures how much the body deviates from a tube-following motion, not slip relative to a surface because there is no surface contact. (F) Center of mass tortuosity $\tau$ (see **Kinematics analysis** in **MATERIALS AND METHODS** for details). In (E) and (F), filled and open bars are for high and low friction, respectively. Error bars show ± 1 s.d. Brackets and/or asterisks represent statistically significant differences between step height treatments (gray) and between surface friction treatments (blue) ($P < 0.05$, ANOVA, Tables S2, 3). (G-J) Top (G, H) and side view (I, J) showing center of mass trajectories of all trials on 15% SVL (G, I) and 30% SVL (H, J) steps. Red solid and blue dashed curves are for high and low friction treatments, respectively. In (G, H), all trajectories are shifted to start at the same lateral location to better show variation.

**Static stability**

For all the step height and surface friction treatments, the snake maintained static stability nearly perfectly during traversal, with its center of mass vertical projection falling within the stability region (Fig. 8A; see example in Movie 2) for nearly 100% of the time during all trials (95% confidence interval: [99.8%, 100.0%]). For all step height and surface friction treatments, stability margin was larger in the pitching direction than in the rolling direction (Fig. 8B, C; $P < 0.0001$, ANOVA, Table S3). On the higher 30% SVL step, regardless of surface friction, pitch stability margin decreased after cantilevering started (during cantilevering, red; after reaching, gray) (Fig. 8A, ii-v) as compared to before cantilevering (Fig. 8A, i; black) (Fig. 8B, black asterisks; $P < 0.05$, ANOVA, Tukey HSD, d.f. = 1, 2, Table S3). On the lower 15% SVL step, pitch stability margin increased with surface friction (Fig. 8B, blue brackets and asterisks; $P < 0.05$, ANOVA, Table S3), whether before or after cantilevering started. On the high friction step, the pitch stability margin after cantilevering started (Fig. 8A, ii-v) decreased with step height (Fig. 8B; $P < 0.05$; ANOVA, Table S3). Surprisingly, on the lower 15% SVL step, roll stability decreased with surface friction (Fig. 8C, blue bracket and asterisk; $P = 0.029$, ANOVA, Table S3).

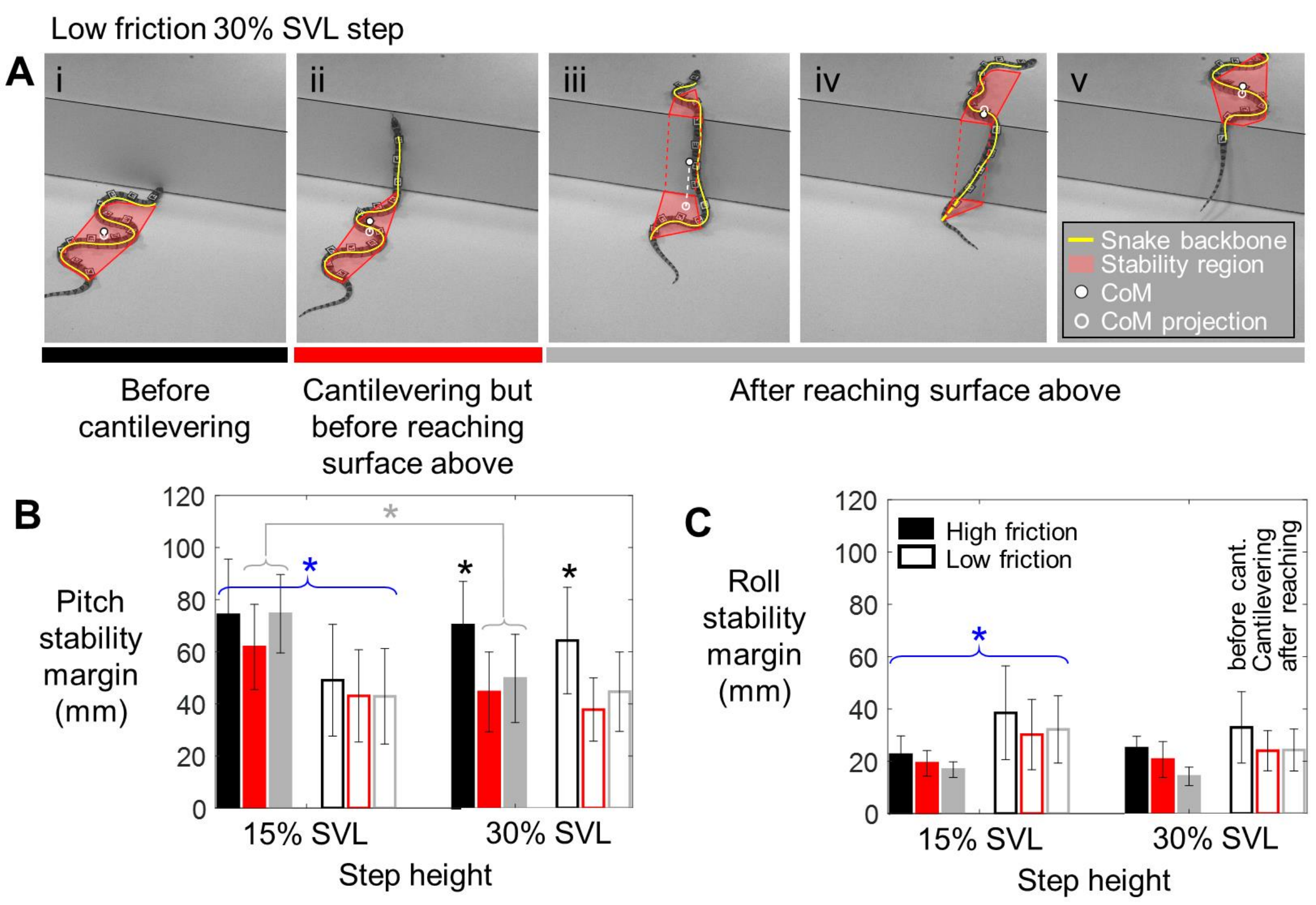

**Fig. 8. Static stability.** (A) Representative oblique view snapshots of snake with static stability region and center of mass overlaid while traversing a low friction 30% SVL step. Yellow curve shows backbone curve, red curves shows boundary of static stability region projected into horizontal surfaces below and above, white solid circle shows center of mass, and white open circle and dashed line show projection of center of mass onto horizontal surfaces below or above. In the fourth snapshot, center of mass is lower than surface above, hence its projection above itself. Note that the non-vertical projection lines (dashed red and dashed white) are an artifact of the oblique view. Colored horizontal bars below snapshots show different stages of traversal: before cantilevering (black, i), during cantilevering before reaching surface above (red, ii), and after reaching surface above (gray, iii-v). (B, C) Static stability margins in the pitching (B) and rolling (C)

directions as a function of step height. Bar colors show stages of traversal defined in (A). Error bars show ± 1 s.d. Brackets and/or asterisks represent statistically significant differences between body sections (black), between step height treatments (gray), and between surface friction treatments (blue) ($P < 0.05$, ANOVA, Table S3).

Only on the most challenging, low friction, higher 30% SVL step did the snake sometimes (8 out of 30 trials) brace a small segment of its body (less than 2 cm length) against the vertical surface before the head reached the surface above the step. On the other three less challenging steps, the snakes did not brace against the vertical surface before the head reach the surface above. In addition, the snake never braced after the head reached the surface above on any of the four steps. Furthermore, after the tail lifted off of the surface below the step, it never braced against the vertical surface for balance (like geckos do with their tails (Jusufi et al., 2008)).

## DISCUSSION

### Partitioned gait conforms to large steps and adjusts to step changes

The snake's partitioned gait, with three body sections that move in different planes (two horizontal and one vertical) and travel down the body, allows it to conform to the large step obstacle throughout traversal (Fig. 9A). Note that to conform here is not necessarily to make contact, because the cantilevering body section moves in the air and rarely contacts the vertical surface to brace against it. Such partitioning of the body into sections, which serve different locomotor functions and are coordinated together to achieve high-level locomotor tasks, has been observed in many arboreal snakes moving on branches or ledges (Jayne and Riley, 2007; Hoefer and Jayne, 2012; Byrnes and Jayne, 2012; Jorgensen and Jayne, 2017; Newman and Jayne, 2018; Lillywhite et al., 2000).

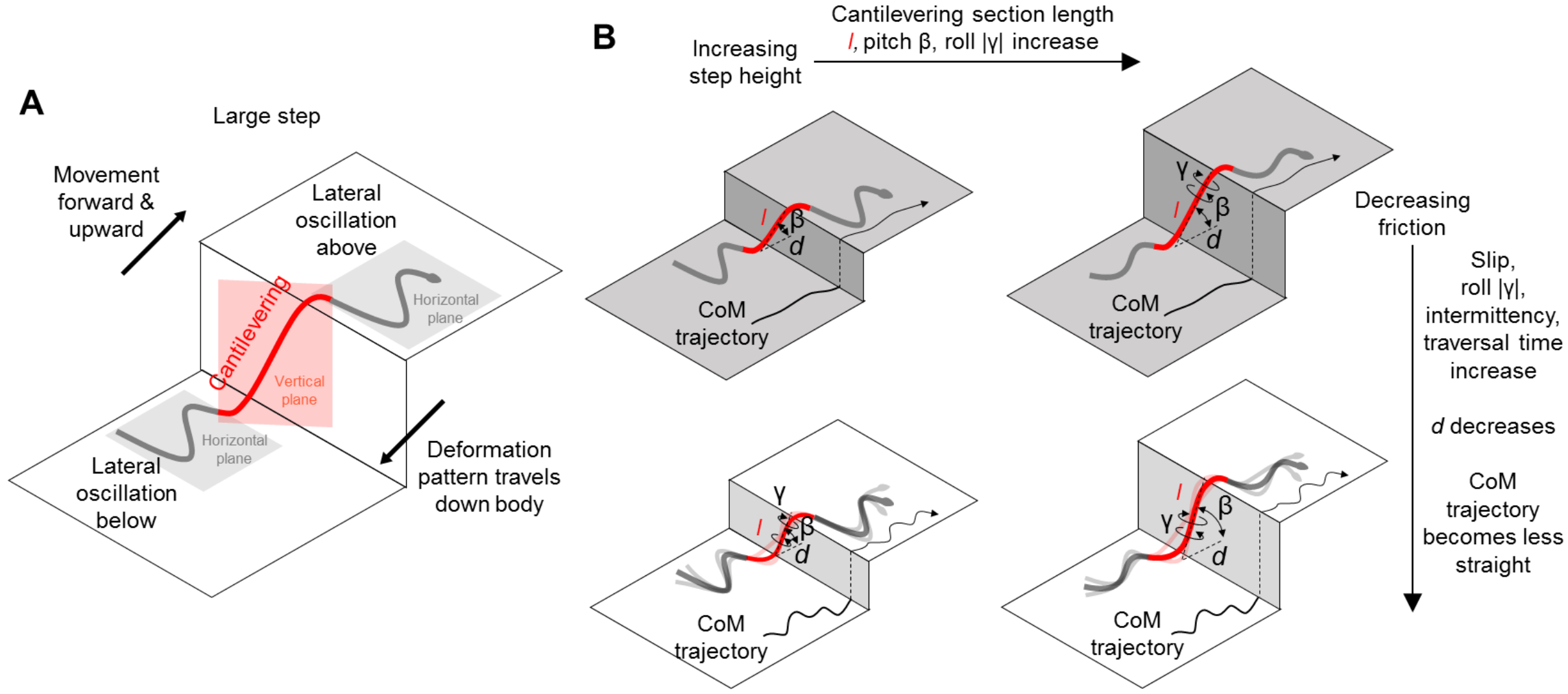

**Fig. 9. Summary of the kingsnake's partitioned gait to traverse a large step of variable height and surface friction.** (A) Snake moves forward and upward to traverse step as lateral oscillation waves travels down its body sections above and below the step to propel (gray), while body section in between cantilevers to bridge the step (red). (B) As step height and surface friction change, snake continues to use partitioned gait, but its kinematics and locomotor performance change in response to variation of terrain properties.

When step properties (step height and surface friction) change, the snake continues to use this partitioned gait, with active adjustments to compensate for, as well as involuntary changes resulting from, terrain variation (Fig. 9B). First, as step height increases, a longer section of the body must be devoted to cantilevering to bridge the step (Fig. 4E-G), which pitches up more (Fig. S3F) but suffers larger rolling or local twisting (at least of the skin) (Fig. S3G). Second, as surface friction decreases, because the snake slips more (Fig. 7A-E), it moves more intermittently (Fig. 5A-E) and progresses more slowly forward and upward (Fig. S3C, E), with a less straight center of mass trajectory (Fig. 7F-J). In addition, as surface friction decreases, the snake initiates cantilevering when it is closer to the step and keeps much of its cantilevering body section closer to step during traversal (Fig. 4A-D, H). Within the snake's body frame, the intermittent self-deformation on the horizontal surfaces most, but far from perfectly, resembles the concertina gait during locomotion on low friction, flat surfaces (Gans, 1975; Jayne, 1986), with body segments alternating between extension and contraction (Fig. 5B, D, Movie 1, parts 3, 4). However, due to frequent, large slipping, body segments do not have the distinct alternating pattern between movement with no slip and no movement at all as seen in a concertina gait. Occasionally, due to large slipping, the intermittent movement resembles slide pushing (Gans, 1984) in that self-deforming body segments progress forward and up the step slowly or do not progress at all (Fig. 5B, D, ii; Fig. 7C, D). The kingsnake's ability to actively adjust its gait in response to changes in step height and surface friction (with concurring involuntary changes) is similar to that of arboreal snakes adjusting the length and orientation of the cantilevering body section in response to changes in branch inclination and diameter (Astley and Jayne, 2007b, Byrnes and Jayne, 2012, Hoefer and Jayne, 2012).

The snake's ability to propagate body sections with distinct movement patterns down its body and to adjust them in response to step changes likely relies on sensory feedback control (Jorgensen and Jayne, 2017). This is because the same feedforward command from the central nervous system (Ijspeert, 2008) that generates body oscillations for the body sections below and above the step can unlikely generate a constant shape within another orthogonal, vertical plane for the body section in between. This is also supported by the observation that arboreal snakes use sensory feedback control during traversal of a large horizontal gap, where the axial muscle activation pattern of its cantilevering body section changes after it has reached across the gap and regains support at both ends, as compared to that during cantilevering before reaching across (Jorgensen and Jayne, 2017). Future experiments using electromyography (Jayne, 1988; Sharpe et al., 2013) and robotic physical models (Astley, 2018; Marvi and Hu, 2012; Marvi et al., 2013) can help reveal how snakes use sensory feedback to control body partitioning to traverse large steps.

**Partitioned gait helps maintain static stability**

Maintaining pitch and roll stability when cantilevering to bridge onto a large step or across a large gap between branches presents a challenge for both arboreal and terrestrial snakes. Although arboreal snakes can grip large asperities such as twigs and secondary branches for stability (Astley and Jayne, 2007a; Jayne and Riley, 2007; Lillywhite et al., 2000), they must also use precise control to laterally distribute body mass equally on branches (Jayne and Herrmann, 2011). By contrast, terrestrial snakes traversing step-like obstacles can use more irregular or asymmetric lateral movements for pitch roll stability, but the lack of large asperities for gripping makes maintaining both pitch and roll stability while cantilevering difficult.

Even more difficulty arises due to lateral or fore-aft perturbations which could cause the cantilevering body section's weight to exert substantial rolling and pitching moments on the body section in contact with the surface below (Astley et al., 2015; Hoefer and Jayne, 2013; Lillywhite et al., 2000), which static stability before the head reaches the surface above (Fig. 8A, ii). This problem is imminent on the low friction steps, where large slipping (Fig. 7C-E) and irregular center of mass movement (Fig. 7C, D, F-J) can induce frequent, large perturbations, especially for the higher step where a longer section of the body must be used for cantilevering. This explains why, when traversing the most challenging low friction, higher 30% SVL step, the snake has to occasionally brace against the vertical surface of the step for stability (8 out of 30 trials).

Except for occasionally bracing its body against the vertical surface (thus generating frictional forces along the vertical surface) on the low friction higher step, the snake mainly counteracts the rolling moment before reaching the surface above by laterally deforming its body section in contact with the surface below to widen the stability region (Fig. 2C-F). Considering this, although snakes similar in body size as the ones studied here (50 cm long (Marvi et al., 2013)) can in theory use a rectilinear-like gait while cantilevering to bridge onto a higher step than using lateral deformation (Gray, 1946; Newman and Jayne, 2018; Marvi et al., 2013), due to its minimal roll stability margin (Byrnes and Jayne, 2012), they would almost certainly tip over with the slightest perturbation, unless the snake promptly braces its body against the vertical surface as soon as cantilevering starts. Thus, there is a tradeoff between maintaining roll stability by lateral body deformation and keeping the body straight to free up a longer body section for cantilevering to reach higher steps.

This function of maintaining roll stability comes at the cost of sacrificing some pitch stability, because lateral body deformation decreases the length of stability region in the fore-aft direction, which counteracts pitching instability (unless the snake braces its body against the vertical surface). Indeed, during traversal of the higher 30% SVL step, pitch stability margin decreased once the snake begins cantilevering but has not reached the surface above (Fig. 8B). In addition, on low friction steps, the snake maintaining much of the cantilevering body section closer to the step (Fig. 4A-F) is likely a response to compensate for higher pitch instability due to frequent, large slipping perturbations. Considering these difficulties, it is remarkable that the snake remained statically stable during cantilevering prior to reaching the surface above the step.

The tradeoff between bridging the step height by body cantilevering and maintaining roll stability by lateral body deformation above and below the step likely limits the highest step that snakes can traverse to well below their maximal vertical cantilevering ability. The kingsnakes in this study only traverse a step of up to 30% SVL (26% of total body length). The maximal vertical cantilevering ability observed in terrestrial snakes is by corn snakes by up to 50% of total body length or 44% SVL (Hoefer and Jayne, 2013; Jayne and Herrmann, 2011), although they are more arboreal than the kingsnakes used in this study. By contrast, arboreal snakes can traverse vertical and horizontal gaps between branches greater than 50% SVL (Hoefer and Jayne, 2013), with the brown tree snake remarkably crossing a vertical gap up to 82% SVL thanks to special musculoskeletal adaptations (Byrnes and Jayne, 2012), and in some cases prehensile tails (Byrnes and Jayne, 2012), along with twigs and secondary branches to use for stability. Considering these special adaptions, arboreal snakes can likely traverse large steps beyond the 30% SVL height that we observed for the generalist kingsnake.

**Partitioned gait may be broadly useful in complex 3-D terrain**

Although only shown here in one species of snake traversing a vertical step connecting two horizontal surfaces, a partitioned gait and the ability to adjust it in response to terrain variation may be a general locomotor adaption of generalist snakes (Gray and Lissmann, 1950; Jayne, 1986) to their diverse and variable habitats consisting of complex 3-D terrain (Li et al., 2015). For example, generalist snakes may use a partitioned gait to cantilever along other directions than vertically upward to traverse two disconnected surfaces that are not horizontal or parallel. We found evidence of this by observing that the kingsnake used a similar, but reversed, partitioned gait to traverse down a large step (Movie 3, part 1) and to traverse a large horizontal gap between two horizontal surfaces (Fu et al., 2018) (Movie 3, part 2). More broadly, to traverse unstructured 3-D terrain like large rocks, felled trees, and rubble, generalist snakes may use multiple small sections of the body to engage some parts of the terrain for support and propulsion and use the body sections in between to bridge across them (Lillywhite et al., 2000). We found evidence of this by observing that the kingsnake partitioned its body into many sections in a similar fashion (alternating between having surface contact and cantilevering) to traverse uneven terrain (similar to (Sponberg and Full, 2008)) (Movie 3, part 3).

**Acknowledgements**

We owe special thanks to Henry Astley for many helpful discussions on snake biomechanics and advice on statistics and animal care. We thank Bruce Jayne, David Hu, Bob Full, Noah Cowan, Dan Goldman, Perrin Schiebel, Jake Socha, Joe Mendelson, and two anonymous reviewers for helpful comments and suggestions; Changxin Yan, Nansong Yi, and Neil McCarter for help with experimental setup and/or preliminary experiments; Qiyuan Fu for taking videos of large step downward traversal and large gap traversal, measuring the number of vertebrae, and help with animal care; Casey Kissel and Mitchel Stover for help with animal euthanization for friction coefficient measurements; and Jin-Seob Kim and Greg Chirikjian for providing initial codes for and technical advice on snake continuous body 3-D kinematics interpolation. All animal experiments were approved by and in compliance with The Johns Hopkins University Animal Care and Use Committee (protocol # RE16A223).

**Competing interests**

The authors declare no competing or financial interests.

## Author contributions

S.W.G designed study, performed locomotion experiments, analyzed data, and wrote the paper; T.W.M. designed study, performed locomotion experiments and friction coefficient measurements, and performed continuous body interpolation and stability region calculations; C.L. designed and supervised the study, defined data analysis, and wrote the paper.

## Funding

This work is funded by a Burroughs Wellcome Fund Career Award at the Scientific Interface and The Johns Hopkins University Whiting School of Engineering start-up funds to C.L. During manuscript revision, S.W.G. is supported by the US Army Research Laboratory.

## Data availability

Data are available from the authors on request.

## Supplementary Figures

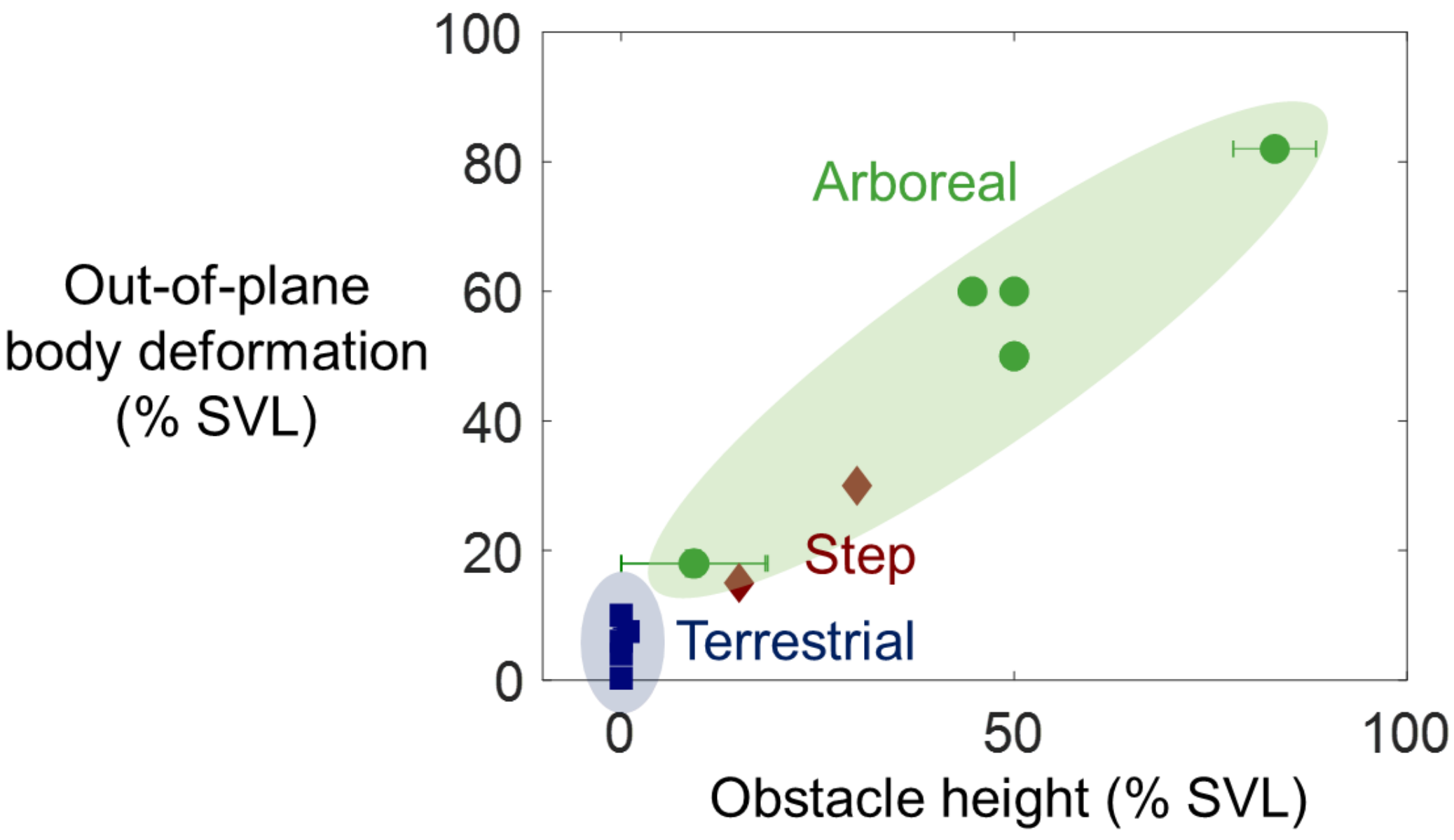

**Figure S1.** Maximal out-of-plane body deformation as a function of obstacle height comparing our study (red diamonds) with previous studies of arboreal (green circles) and terrestrial (blue squares) snake locomotion (see Table S1).

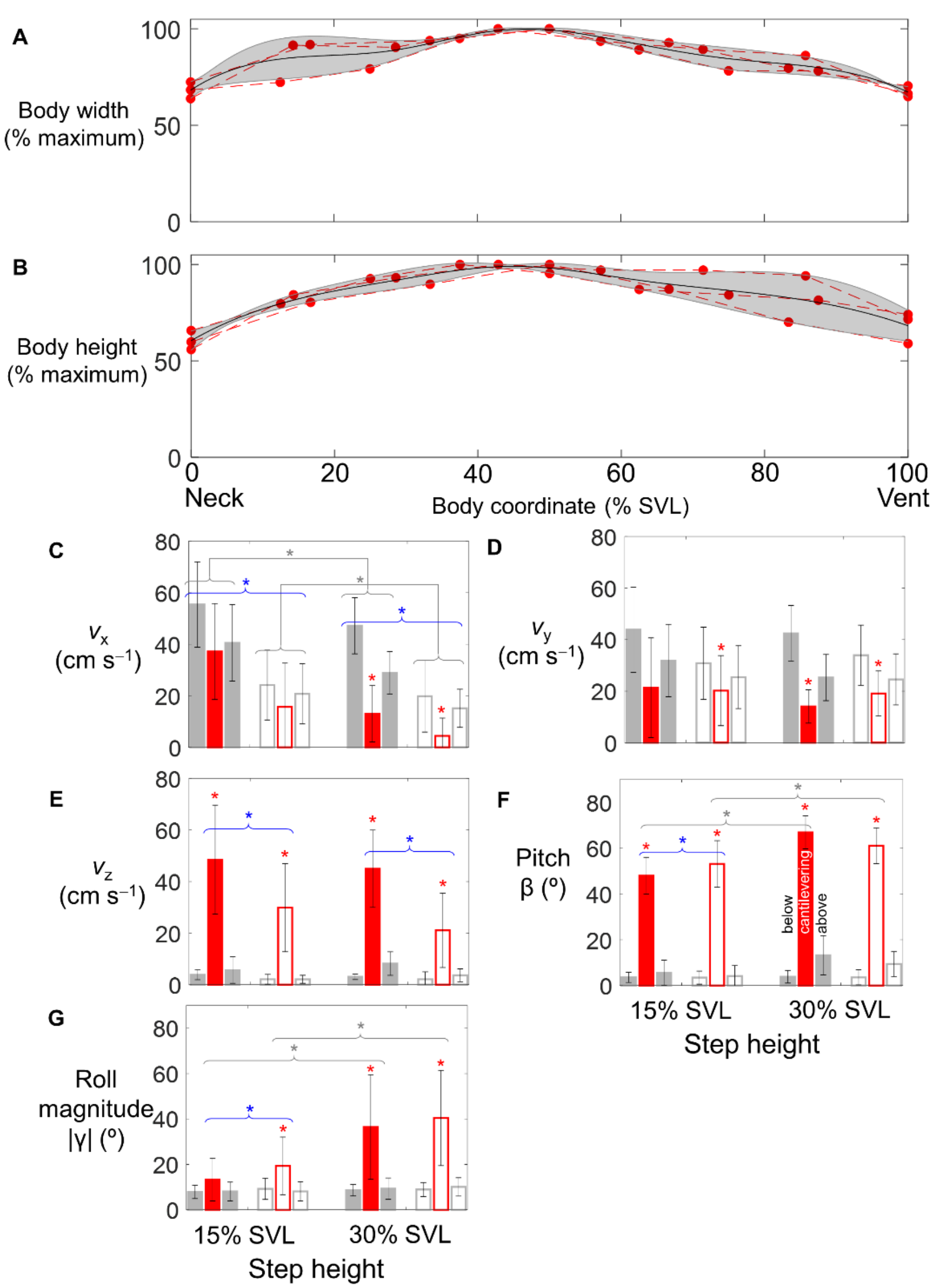

**Figure S2.** Body width (A) and height (B) along the body. Red markers and dotted lines show measurement points for each individual. Black curves and shaded area show mean ± 1 s.d. Because we could not measure at the exact same body coordinates for each snake, we interpolated measurements along the body. Forward (C), lateral (D), and vertical (E) speeds as a function of step height. Body pitch $\beta$ (F) and roll magnitude $|\gamma|$ (considering lateral symmetry) (G) as a function of step height. Filled and open bars are for high and low friction treatments, respectively. For each treatment, two gray bars are for body section below (left) and above (right) step, and red bar in between is for cantilevering body section. Error bars show ± 1 s.d. Brackets and/or asterisks represent statistically significant differences between body sections (black), between step height treatments (gray), and between surface friction treatments (blue) ($P < 0.05$, ANOVA, Table S3). Connected brackets represent a significant difference across treatments for all body sections ($P < 0.05$, ANOVA, Table S3).

## Supplementary Tables

**Table S1.** Maximal out-of-plane movement in previous studies of snake locomotion. Starred (*) studies observed substantial out-of-plane body deformation (> 10% snout-vent length). Many other early studies investigated snake movement on flat surfaces but no out-of-plane data were available.

| Study | Terrain | Species | SVL (cm) | Obstacle height (cm) | Maximal out-of-plane movement | |
|---|---|---|---|---|---|---|
| | | | | | (cm) | (% SVL) |
| (Marvi and Hu, 2012) | Narrow channel | Corn snake (*Elaphe guttata*) | 61 ± 4 | 0 | ≈ 0.2 | 0.30 ± 0.04 |
| (Marvi et al., 2014) | Sandy slope | Sidewinder rattlesnake (*Crotalus cerastes*) | 48 ± 6 | 0 | 3 | 6 ± 2 |
| (Jafari et al., 2014) | Air gliding | Paradise tree snake (*Chrysopelea paradise*) | 60.3, 74.0 | N.A. | 6, 7 | 10, 10 |
| (Jayne and Riley, 2007) | Tree branches | Brown tree snake (*Boiga irregularis*) | 43-188 | 0 | 2.2-6.4 | 3-5 |
| (Byrnes and Jayne, 2010) | Tree branches | Boa constrictor (*Boa constrictor*) | 66-70 | 0.3-0.9 | ~5 | 7-8 |
| *(Jayne and Herrmann, 2011) | Tree branches | Boa constrictor (*Boa constrictor*) | 60.0 ± 0.6 | 0.2-10.8 | 10.8 | 18 ± 2 |
| | | Corn snake (*Pantherophis guttatus*) | 59.0 ± 0.6 | 0.2-10.8 | 10.8 | 18 ± 2 |
| *(Byrnes and Jayne, 2012) | Tree branches | Brown tree snake (*Boiga irregularis*) | 90-102 | 74-84 | 74-84 | 82 |
| *(Hoefer and Jayne, 2013) | Tree branches | Boa constrictor (*Boa constrictor*) | 84 | 42 | 42 | 50 |
| | | Corn snake (*Pantherophis guttatus*) | 68 | 34 | 34 | 50 |
| | | Brown tree snake (*Boiga irregularis*) | 84 | 42 | 50 | 60 |
| | | Brown tree snake (*Boiga irregularis*) | 68 | 34 | 41 | 60 |
| Our study | Step | Kingsnake (*Lampropeltis Mexicana*) | 34.6 ± 0.4 | 5-10.5 | 10.5 | 30 |

**Table S2. Results from ANOVAs testing the effects of step height and surface friction on traversal performance.** *F* ratios and *P* values are shown as *F*(*P*). *N* = 3 individuals and *n* = 120 trials in total (10 trials for each individual each treatment). Individual is set as a random, crossed factor to account for individual variation. Results for the random individual factor are not shown for simplicity. See **Statistics** in **MATERIALS AND METHODS** for detail.

| Effect | Degree of freedom | Dependent variables | | | | | | |
|---|---|---|---|---|---|---|---|---|
| | | Traversal time | Traversal speed $v_{CoM}$ | Velocity intermittency | Cantilever length $l$ | Tortuosity $\tau$ | Head lift-off distance | Body-lift-off distance |
| Height | 1, 2 | 12.6 (< 0.001) | 5.54 (0.1428) | 42.0 (0.023) | 198.1 (0.005) | 6.9 (0.119) | 1.03 (0.312) | 4.54 (0.0353) |
| Friction | 1, 2 | 33.5 (< 0.001) | 18696.4 (< 0.001) | 42.5 (< 0.001) | 2.8 (0.234) | 12.7 (0.0704) | 106.8 (< 0.001) | 35.6 (< 0.001) |
| Height × Friction | 1, 2 | 0.036 (0.85) | 1.9 (0.298) | 13.1 (0.643) | 5.74 (0.139) | 6.9 (0.1199) | 3.56 (0.0620) | 5.3 (0.0232) |

**Table S3. Results from ANOVAs testing the effects of step height, surface friction, and body section (below step, cantilevering, and above step) on traversal kinematics.** *F* ratios and *P* values are shown as *F*(*P*). $N = 3$ individuals and $n = 120$ trials in total (10 trials for each individual each treatment). Individual is set as a random, crossed factor to account for individual variation. Results for the random individual factor are not shown for simplicity. For pitch and roll stability margins, stage of traversal is used as a fixed effect; for the remaining measurements, body section is used as a fixed effect; both were indicated by S in the effect column. See **Statistics** in **MATERIALS AND METHODS** for detail.

| Effect | Degree of freedom | Dependent variables | | | | | | | | | |
|---|---|---|---|---|---|---|---|---|---|---|---|
| | | In-plane length | Local curvature κ | Slip angle φ | Forward speed $v_x$ | Lateral speed $v_y$ | Vertical speed $v_z$ | Pitch β | Roll magnitude \|γ\| | Pitch stability margin | Roll stability margin |
| Height (H) | 1, 2 | 1.0 (0.43) | 39.6 (0.024) | 47.0 (0.021) | 9.6 (0.091) | 0.46 (0.57) | 4.5 (0.17) | 277.3 (0.004) | 33.4 (0.029) | 2.0 (0.2954) | 6.0 (0.13) |
| Friction (F) | 1, 2 | 110.3 (0.008) | 14.3 (0.063) | 156.1 (0.006) | 341.7 (0.0029) | 43.2 (0.022) | 2318.5 (< 0.001) | 1.2 (0.39) | 1.0 (0.41) | 19.0 (0.048) | 32.8 (0.029) |
| Section/Stage (S) | 2, 4 | 51.1 (0.001) | 200.7 (< 0.001) | 1.3 (0.36) | 155.6 (< 0.001) | 177.3 (< 0.001) | 24.4 (0.006) | 407.5 (< 0.001) | 19.5 (0.009) | 13.8 (0.0157) | 16.0 (0.0119) |
| H × F | 1, 2 | 22.1 (0.041) | 2.5 (0.26) | 38.9 (0.024) | 9.1 (0.091) | 7.9 (0.11) | 0.84 (0.46) | 41.8 (0.023) | 0.53 (0.54) | 12.1 (0.073) | 39.1 (0.022) |
| F × S | 2, 4 | 70.0 (< 0.001) | 0.122 (0.89) | 1.1 (0.36) | 8.8 (0.034) | 19.0 (0.009) | 791.7 (< 0.001) | 1.4 (0.34) | 1.0 (0.54) | 1.2 (0.38) | 21.5 (0.004) |
| H × S | 2, 4 | 10.5 (0.025) | 69.7 (< 0.001) | 7.5 (0.044) | 11.8 (0.021) | 16.7 (0.011) | 17.6 (0.010) | 98.3 (< 0.001) | 16.5 (0.011) | 79.1 (< 0.001) | 3.6 (0.12) |
| H × F × S | 2, 4 | 9.5 (0.030) | 9.0 (0.033) | 1.2 (0.40) | 1.2 (0.38) | 0.103 (0.905) | 1.1 (0.42) | 20.1 (0.008) | 5.1 (0.078) | 7.1 (0.043) | 0.15 (0.86) |

**Table S4. Results from ANOVAs testing the effects of step height, surface friction, and body section (below and above step only, no cantilevering) on traversal kinematics.** *F* ratios and *P* values are shown as *F*(*P*). *N* = 3 individuals and *n* = 120 trials in total (10 trials for each individual each treatment). Individual is set as a random, crossed factor to account for individual variation. Results for the random individual factor are not shown for simplicity. See **Statistics** in **MATERIALS AND METHODS** for detail.

| Effect | Degree of freedom | Dependent variables | | | |
|---|---|---|---|---|---|
| | | Lateral oscillation amplitude *A* | Lateral oscillation frequency *f* | Lateral oscillation wavelength λ | Lateral oscillation wavenumber σ |
| Height (H) | 1, 2 | 167.5 (0.005) | 0.01 (0.95) | 11.2 (0.078) | 5.8 (0.137) |
| Friction (F) | 1, 2 | 5.0 (0.16) | 10.7 (0.082) | 24.8 (0.034) | 13.3 (0.067) |
| Section (S) | 1, 2 | 1.9 (0.30) | 0.31 (0.63) | 47.3 (0.020) | 8.9 (0.096) |
| H × F | 1, 2 | 0.17 (0.72) | 10.1 (0.084) | 0.23 (0.68) | 1.7 (0.32) |
| F × S | 1, 2 | 0.06 (0.95) | 1.6 (0.34) | 0.17 (0.722) | 0.23 (0.68) |
| H × S | 1, 2 | 2.3 (0.27) | 0.77 (0.48) | 1.1 (0.40) | 1.6 (0.33) |
| H × F × S | 1, 2 | 0.25 (0.67) | 0.46 (0.57) | 1.4 (0.35) | 0.16 (0.73) |

## Supplementary Movies

Movie 1. A kingsnake partitions its body to traverse a large step obstacle. The four video clips are for a high friction 15% SVL step (part 1), a high friction 30% SVL step (part 2), a low friction 15% SVL step (part 3), and a low friction 30% SVL step (part 4), respectively.

https://www.youtube.com/watch?v=RsT4IHM366U

Movie 2. A kingsnake maintains static stability throughout the traversal of a large step using body partitioning.

https://www.youtube.com/watch?v=Opi3xO0Tvok

Move 3. A kingsnake partitions its body to traverse other complex 3-D terrain. The three video clips are for traversing a large step downward (part 1), a large gap (part 2), and rough terrain (part 3), respectively. Parts 1 and 2 credit: Qiyuan Fu.

https://www.youtube.com/watch?v=3ZUHB9Qtir0